\documentclass[a4paper,11pt]{article}
\usepackage[latin1]{inputenc}
\usepackage[T1]{fontenc}
\usepackage[english]{babel}
\usepackage{amssymb}
\usepackage{amsfonts, amsmath}
\newtheorem{Prop}{Proposition}[section]
\newtheorem{Lem}{Lemma}[section]
\newtheorem{Rem}{Remark}[section]
\title{\textbf{Affine parts of abelian surfaces as complete intersection of three quartics}}
\author{\textbf{A. Lesfari}
\\\emph{Department of Mathematics}
\\\emph{Faculty of Sciences}
\\\emph{University of Choua\"{i}b Doukkali}
\\\emph{B.P. 20, El-Jadida, Morocco}.
\\\emph{E. mail} : Lesfariahmed@yahoo.fr, Lesfari@ucd.ac.ma}
\date{}
\begin{document}
\maketitle
\begin{abstract}
We consider an integrable system in five unknowns having three
quartics invariants. We show that the complex affine variety
defined by putting these invariants equal to generic constants,
completes into an abelian surface; the jacobian of a genus two
hyperelliptic curve. This system is algebraic completely
integrable and it can be integrated in genus two hyperelliptic
functions.

\end{abstract}
\section{Introduction}
The problem of finding and integrating hamiltonian systems, has
attracted a considerable amount of attention in recent years.
Beside the fact that many integrable hamiltonian systems have been
on the subject of powerful and beautiful theories of mathematics,
another motivation for its study is: the concepts of integrability
have been applied to an increasing number of physical systems,
biological phenomena, population dynamics, chemical rate
equations, to mention only a few. However, it seems still hopeless
to describe or even to recognize with any facility, those
hamiltonian systems which are integrable, though they are quite
exceptional. In this paper, we shall be concerned with finite
dimensional algebraic completely integrable systems. A dynamical
system is algebraic completely integrable if it can be linearized
on a complex algebraic torus $\mathbb{C}^{n}/lattice$ (=abelian
variety). The invariants (often called first integrals or
constants) of the motion are polynomials and the phase space
coordinates (or some algebraic functions of these) restricted to a
complex invariant variety defined by putting these invariants
equals to generic constants, are meromorphic functions on an
abelian variety. Moreover, in the coordinates of this abelian
variety, the flows (run with complex time) generated by the
constants of the motion are straight lines. However, besides the
fact that many hamiltonian completely integrable systems posses
this structure, another motivation for its study which sounds more
modern is: algebraic completely integrable systems come up
systematically whenever you study the isospectral deformation of
some linear operator containing a rational indeterminate.
Therefore there are hidden symmetries which have a group
theoretical foundation. The concept of algebraic complete
integrability is quite effective in small dimensions and has the
advantage to lead to global results, unlike the existing criteria
for real analytic integrability, which, at this stage are
perturbation results. In fact, the overwhelming majority of
dynamical systems, hamiltonian or not, are non-integrable and
possess regimes of chaotic behavior in phase space. In the present
paper, we discuss an interesting interaction between complex
algebraic geometry and dynamical systems. we present an integrable
system in five unknowns having three quartics invariants. This
system is algebraic completely integrable in $\mathbb{C}^5,$ it
can be integrated in genus 2 hyperelliptic functions. We show that
the complex affine variety B(3) defined by putting these
invariants equal to generic constants, is a double cover of a
Kummer surface and the system (1) can be integrated in genus 2
hyperelliptic functions. We make a careful study of the algebraic
geometric aspect of the affine variety B(3) of the system (1). We
find via the Painlevé analysis the principal balances of the
hamiltonian field defined by the hamiltonian. To be more precise,
we show that the system (1) possesses Laurent series solutions in
$t$, which depend on 4 free parameters : $\alpha, \beta, \gamma$
and $\theta.$ These meromorphic solutions restricted to the
surface $B$(3) are parameterized by two isomorphic smooth
hyperelliptic curves $\mathcal{H}_{\varepsilon=\pm i}$(8) of genus
2 that intersect in only one point at which they are tangent to
each other. The affine variety $B$(3) is embedded into
$\mathbb{P}^{15}$ and completes into an abelian variety
$\widetilde{B}$ (the jacobian of a genus 2 curve) by adjoining a
divisor $\mathcal{D}=\mathcal{H}_i+\mathcal{H}_{-i}.$ The latter
has geometric genus 5 and $\mathcal{S}=2\mathcal{D}$ (very ample)
has genus 17. The flow (1) evolves on $\widetilde{B}$ and is
tangent to each hyperelliptic curve $\mathcal{H}_\varepsilon$ at
the point of tangency between them. Consequently, the system (1)
is algebraic integrable.

Abelian varieties, very heavily studied by algebraic geometers,
enjoy certain algebraic properties which can then be translated
into differential equations and their Laurent solutions. Among the
results presented in this paper, there is an explicit calculation
of invariants for a hamiltonian system which cut out an open set
in an abelian variety and various curves related to this system
are given explicitly. The integrable dynamical system presented
here is interesting, particular to experts of abelian varieties
who may want to see explicit examples of a correspondence for
varieties defined by different curves.\\

\section{A five-dimensional integrable system}

Let us consider the following system of five differential
equations in the unknowns $z_1,\ldots,z_5$ :
\begin{eqnarray}
\dot{z}_1&=&2z_4,\nonumber\\
\dot{z}_2&=&z_3,\nonumber\\
\dot{z}_3&=&-4a z_{2}-6z_{1}z_{2}-16z_{2}^{3} ,\\
\dot{z}_4&=&-az_{1}-z_{1}^{2}-8z_{1}z_{2}^{2}+z_{5},\nonumber\\
\dot{z}_5&=&-8z_{2}^{2}z_{4}-2az_{4}-2z_{1}z_{4}+4z_{1}z_{2}z_{3},\nonumber
\end{eqnarray}
where the dot denotes differentiation with respect to the time
$t$.
\begin{Prop}
The system (1) possesses three quartic invariants and is
completely integrable in the sense of Liouville. The complex
affine variety B(3) defined by putting these invariants equal to
generic constants, is a double cover of a Kummer surface(4) and
the system (1) can be integrated in genus 2 hyperelliptic
functions.
\end{Prop}
\emph{Proof}. The following three quartics are constants of motion
for this system
\begin{eqnarray}
F_1&=&\frac{1}{2}z_{5}+2z_{1}z_{2}^{2}+\frac{1}{2}z_{3}^{2}
+\frac{1}{2}az_{1}+2az_{2}^{2}+\frac{1}{4}z_{1}^{2}+4z_{2}^{4},\nonumber\\
F_2&=&az_{1}z_{2}+z_{1}^{2}z_{2}+4z_{1}z_{2}^{3}-z_{2}z_{5}+z_{3}z_{4} ,\\
F_3&=&z_{1}z_{5}-2z_{1}^{2}z_{2}^{2}-z_{4}^{2}.\nonumber
\end{eqnarray}
The system (1) can be written as a hamiltonian vector field
$$\dot{z}=J\frac{\partial
H}{\partial z},\quad z=(z_{1},z_{2},z_{3},z_{4},z_{5})^\top,$$
where $H=F_{1}.$ The hamiltonian structure is defined by the
Poisson bracket
$$\left\{ F,H\right\} =\left\langle \frac{\partial F}{\partial z},
J\frac{\partial H}{\partial z}\right\rangle
=\sum_{k,l=1}^{5}J_{kl}\frac{\partial F}{\partial
z_{k}}\frac{\partial H}{\partial z_{l}},$$ where
$$\frac{\partial H}{\partial
z}=(\frac{\partial H}{\partial z_{1}},\frac{\partial H}{\partial
z_{2}},\frac{\partial H}{\partial z_{3}},\frac{\partial
H}{\partial z_{4}},\frac{\partial H}{\partial z_{5}})^\top,$$ and
$$J=\left[\begin{array}{ccccc}
0&0&0&2z_1&4z_4\\
0&0&1&0&0\\
0&-1&0&0&-4z_1z_2\\
-2z_1&0&0&0&2z_5-8z_1z_{2}^{2}\\
-4z_4&0&4z_1z_2&-2z_5+8z_1z_{2}^{2}&0
\end{array}\right],$$
is a skew-symmetric matrix for which the corresponding Poisson
bracket satisfies the Jacobi identities. The second flow commuting
with the first is regulated by the equations
$$\dot{z}=J\frac{\partial F_{2}}{\partial z},\quad
z=(z_{1},z_{2},z_{3},z_{4},z_{5})^\top,$$ and is written
explicitly as
\begin{eqnarray}
\dot{z}_1&=&2z_{1}z_{3}-4z_{2}z_{4},\nonumber\\
\dot{z}_2&=&z_4,\nonumber\\
\dot{z}_3&=&z_{5}-8z_{1}z_{2}^{2}-az_{1}-z_{1}^{2} ,\nonumber\\
\dot{z}_4&=&-2az_{1}z_{2}-4z_{1}^{2}z_{2}-2z_{2}z_{5},\nonumber\\
\dot{z}_5&=&-4az_{2}z_{4}-4z_{1}z_{2}z_{4}-16z_{2}^{3}z_{4}-2z_{3}z_{5}
+8z_{1}z_{2}^{2}z_{3}.\nonumber
\end{eqnarray}
These vector fields are in involution, i.e.,
$$\{F_1,F_2\}=\langle \frac{\partial F_{1}}{\partial z},J\frac{\partial F_{2}}{\partial
z}\rangle=0,$$ and the remaining one is casimir, i.e.,
$$J\frac{\partial F_{3}}{\partial z}=0.$$
Let B be the complex affine variety defined by
\begin{equation}\label{eqn:euler}
B=\bigcap_{k=1}^{2}\{z:F_k(z)=c_k\}\subset\mathbb{C}^5.
\end{equation}
Since $B$ is the fibre of a morphism from $\mathbb{C}^5$ to
$\mathbb{C}^3$ over $( c_{1},c_{2},c_{3}) \in \mathbb{C}^{3}$, for
almost all $c_1, c_2, c_3,$ therefore $B$ is a smooth affine
surface. Note that
$$\sigma :(z_{1},z_{2},z_{3},z_{4},z_{5}) \longmapsto (z_{1},z_{2},-z_{3},-z_{4},z_{5}),$$
is an involution on $B$. The quotient $B/\sigma$ is a Kummer
surface defined by
\begin{equation}\label{eqn:euler}
p\left( z_{1},z_{2}\right) z_{5}^{2}+q\left( z_{1},z_{2}\right)
z_{5}+r\left( z_{1},z_{2}\right) =0,
\end{equation}
where
\begin{eqnarray}
p\left( z_{1},z_{2}\right)& =&z_{2}^{2}+z_{1},\nonumber\\
q\left( z_{1},z_{2}\right)
&=&\frac{1}{2}z_{1}^{3}+2az_{1}z_{2}^{2}
+az_{1}^{2}-2c_{1}z_{1}+2c_{2}z_{2}-c_{3},\nonumber\\
r\left( z_{1},z_{2}\right)& =&-8c_{3}z_{2}^{4}+\left( a
^{2}+4c_{1}\right) z_{1}^{2}z_{2}^{2}-8c_{2}z_{1}z_{2}^{3}-
2c_{2}z_{1}^{2}z_{2}-4c_{3}z_{1}z_{2}^{2}\nonumber\\\nonumber\\
&&-\frac{1}{2}c_{3}z_{1}^{2}-4a c_{3}z_{2}^{2}-2a
c_{2}z_{1}z_{2}-a c_{3}z_{1}+c_{2}^{2}+2c_{1}c_{3}.\nonumber
\end{eqnarray}
Using $F_1=c_1$ (2), we have
$$z_{5}=2c_{1}-4z_{1}z_{2}^{2}-z_{3}^{2}-az_{1}-4az_{2}^{2}-
\frac{1}{2}z_{1}^{2}-8z_{2}^{4},$$ and substituting this into
$F_2=c_2,$ $F_3=c_3,$ (2) yields
\begin{eqnarray}
&&2az_{1}z_{2}+\frac{3}{2}z_{1}^{2}z_{2}+8z_{1}z_{2}^{3}-2c_{1}z_{2}
+z_{2}z_{3}^{2}+\allowbreak
4az_{2}^{3}+8z_{2}^{5}+z_{3}z_{4}=c_{2},\nonumber\\
&&2c_{1}z_{1}-6z_{1}^{2}z_{2}^{2}-z_{1}z_{3}^{2}-az_{1}^{2}-4az_{1}z_{2}^{2}-\allowbreak
\frac{1}{2}z_{1}^{3}-8z_{1}z_{2}^{4}-z_{4}^{2}=c_{3}.
\end{eqnarray}
We introduce two coordinates $s _{1}, s _{2}$ as follows
\begin{eqnarray}
z_{1}&=&-4s_{1}s_{2},\nonumber\\
z_{2}&=&s_{1}+s_{2},\nonumber\\
z_{3}&=&\dot{s}_{1}+\dot{s}_{2},\nonumber\\
z_{4}&=&-2\left(
\dot{s}_{1}s_{2}+s_{1}\dot{s}_{2}\right).\nonumber
\end{eqnarray}
Upon substituting this parametrization, (5) turns into
\begin{eqnarray}
&&\left( s_{1}-s_{2}\right) \left(
(\dot{s}_{1})^{2}-(\dot{s}_{2})^{2}\right)+8\left(s_{1}+s_{2}\right)
\left(s_{1}^{4}+s_{2}^{4}+s_{1}^{2}s_{2}^{2}\right)\nonumber\\
&&+4a\left(s_{1}+s_{2}\right) \left(s_{1}^{2}+s_{2}^{2}\right)
-2c_{1}\left(s_{1}+s_{2}\right)-c_{2}=0,\nonumber\\
&&\left(s_{1}-s_{2}\right)
\left(s_{2}(\dot{s}_{1})^{2}-s_{1}(\dot{s}_{2})^{2}\right)+32s_{1}s_{2}\left(
s_{1}^{4}+s_{2}^{4}+s_{1}^{2}s_{2}^{2}\right) \nonumber\\
&&+32s_{1}^{2}s_{2}^{2}\left(s_{1}^{2}+s_{2}^{2}\right)+16as_{1}s
_{2}\left(s_{1}^{2}+s_{2}^{2}\right) +\allowbreak 16as
_{1}^{2}s_{2}^{2}-8c_{1}s_{1}s_{2}-c_{3}=0.\nonumber
\end{eqnarray}
These equations are solved linearly for $\dot{s}_{1}^{2}$ and
$\dot{s}_{2}^{2}$ as
\begin{eqnarray}
(\dot{s}_{1})^{2}&=&\frac{-32s_{1}^{6}-16a s_{1}^{4}+8c_{1}s
_{1}^{2}+4c_{2}s _{1}-c_{3}}{4\left(s
_{2}-s _{1}\right) ^{2}},\\
(\dot{s}_{2})^{2}&=&\frac{-32s_{2}^{6}-16a s _{2}^{4}+8c_{1}s
_{2}^{2}+4c_{2}s_{2}-c_{3}}{4\left(s _{2}-s _{1}\right)
^{2}},\nonumber
\end{eqnarray}
and can be integrated by means of the Abel transformation
$\mathcal{H}\longrightarrow Jac(\mathcal{H}),$ where the
hyperelliptic curve $\mathcal{H}$ of genus 2 is given by an
equation $$w^2=-32s^{6}-16a s^{4}+8c_{1}s ^{2}+4c_{2}s-c_{3}.$$
Consequently, the equations (1) are integrated in terms of genus 2
hyperelliptic functions. This establishes the proposition.

\section{Laurent series solutions and algebraic curves}

The invariant variety $B$(3) is a smooth affine surface for
generic values of $c_{1},c_{2}$ and $c_{3}.$ So, the question I
address is how does one find the compactification of $B$ into an
abelian surface? Following the methods in Adler-van Moerbeke [1],
the idea of the direct proof is closely related to the geometric
spirit of the (real) Arnold-Liouville theorem $\left[8\right] $.
Namely, a compact complex $n$-dimensional variety on which there
exist $n$ holomorphic commuting vector fields which
are independent at every point is analytically isomorphic to a $n$%
-dimensional complex torus $\mathbb{C}^{n}/Lattice$ and the
complex flows generated by the vector fields are straight lines on
this complex torus. Now, the main problem will be to complete
$B\left(3\right) $\ into a non singular compact complex algebraic
variety $\widetilde{B}=B\cup \mathcal{D}$ in such a way that the
vector fields $X_{F_1}$ and $X_{F_2}$ generated respectively by
$F_1$ and $F_2$, extend holomorphically along a divisor
$\mathcal{D}$ and remain independent there. If this is possible,
$\widetilde{B}$ is an algebraic complex torus (an abelian variety)
and the coordinates $z_{1},\ldots,z_{5}$ restricted to $B$ are
abelian functions. A naive guess would be to take the natural
compactification $\overline{B}$ of $B$ by projectivizing the
equations: $$\overline{B}=\bigcap_{k=1}^{3}\{F_{k}(Z)=c_kZ_0^4\}
\subset \mathbb{P}^{5}.$$
Indeed, this can never work for a general reason : an abelian variety $%
\widetilde{B}$ of dimension bigger or equal than two is never a
complete smooth intersection, that is it can never be described in some projective space $%
\mathbb{P}^{n}$ by $n$--dim $\widetilde{B}$ global polynomial
homogeneous equations. In other words, if $B$ is to be the affine
part of an abelian surface, $\overline{B}$ must have a singularity
somewhere along the locus at infinity $\overline{B}\cap \left\{
Z_{0}=0\right\} .$ In fact, we shall show that the existence of
meromorphic solutions to the differential equations (1) depending
on 4 free parameters can be used to manufacture the tori, without
ever going through the delicate procedure of blowing up and down.
Information about the tori can then be gathered from the divisor.

\begin{Prop}
The system (1) possesses Laurent series solutions which depend on
4 free parameters : $\alpha, \beta, \gamma $ and $\theta.$ These
meromorphic solutions restricted to the surface B(3) are
parameterized by two isomorphic smooth hyperelliptic curves
$\mathcal{H}_{\varepsilon=\pm i}$(8) of genus 2.
\end{Prop}
\emph{Proof}. The first fact to observe is that if the system is
to have Laurent solutions depending on 4 free parameters, the
Laurent decomposition of such asymptotic solutions must have the
following form
\begin{eqnarray}
z_{1}&=&\frac{1}{t}(z_{1}^{(0)}+z_{1}^{(1)}t+z_{1}^{(2)}t^{2}
+z_{1}^{(3)}t^{3}+z_{1}^{(4)}t^{4}+\cdots ),\nonumber\\
z_{2}&=&\frac{1}{t}(z_{2}^{(0)}+z_{2}^{(1)}t+z_{2}^{(2)}t^{2}
+z_{2}^{(3)}t^{3}+z_{2}^{(4)}t^{4}+\cdots ),\nonumber\\
z_{3}&=&\frac{1}{t^{2}}(-z_{2}^{(0)}+z_{2}^{(2)}t^{2}+2z_{2}^{(3)}t^{3}
+3z_{2}^{(4)}t^{4}+\cdots ), \nonumber\\
z_{4}&=&\frac{1}{2t^{2}}(-z_{1}^{(0)}+z_{1}^{(2)}t^{2}+2z_{1}^{(3t)
}t^{3}+3z_{1}^{(4) }t^{4}+\cdots ), \nonumber\\
z_{5}&=&\frac{1}{t^{3}}(z_{5}^{(0)}+z_{5}^{(1)}t+z_{5}^{(2)}t^{2}
+z_{5}^{(3)}t^{3}+z_{5}^{(4)}t^{4}+\cdots ). \nonumber
\end{eqnarray}
Putting these expansions into
\begin{eqnarray}
\ddot{z}_{1}&=&-2az_{1}-2z_{1}^{2}-16z_{1}z_{2}^{2}+2z_{5},\nonumber\\
\ddot{z}_{2}&=&-4az_{2}-6z_{1}z_{2}-16z_{2}^{3},\nonumber\\
\dot{z}_{5}&=&-8z_{2}^{2}z_{4}-2az_{4}-2z_{1}z_{4}+4z_{1}z_{2}z_{3},\nonumber
\end{eqnarray}
deduced from (1), solving inductively for the
$z_{k}^{(j)}(k=1,2,5),$ one finds at the $0^{th}$ step (resp.
$2^{th}$ step) a free parameter $\alpha$ (resp. $\beta$) and the
two remaining ones $\gamma,\theta$ at the $4^{th}$ step. More
precisely, we have
\begin{eqnarray}
z_{1}&=&\frac{1}{t}(\alpha -\alpha ^{2}t+\beta t^{2}
+\frac{1}{6}\alpha (3\beta -9\alpha ^{3}+4a\alpha)t^{3}+\gamma t^{4}+\cdots ),\nonumber\\
z_{2}&=&\frac{\varepsilon \sqrt{2}}{4t}(1+\alpha
t+\frac{1}{3}(-3\alpha ^{2}+2a)t^{2}+\frac{1}{2}(3\beta -\alpha
^{3})t^{3}-2\varepsilon \sqrt{2}\theta t^{4}+\cdots), \nonumber\\
z_{3}&=&\frac{\varepsilon
\sqrt{2}}{4t^{2}}(-1+\frac{1}{3}(-3\alpha ^{2}
+2a)t^{2}+(3\beta -\alpha ^{3})t^{3}-6\varepsilon \sqrt{2}\theta t^{4}+\cdots),\\
z_{4}&=&\frac{1}{2t^{2}}(-\alpha +\beta t^{2}+\frac{1}{3}\alpha
(3\beta -9\alpha ^{3}
+4a\alpha)t^{3}+3\gamma t^{4}+\cdots), \nonumber\\
z_{5}&=&\frac{1}{t}(-\frac{1}{3}a \alpha +\alpha ^{3}-\beta
+( 3\alpha ^{4}-a\alpha ^{2} -3\alpha \beta)t \nonumber\\
&&+( 4\varepsilon \sqrt{2}\alpha \theta +2\gamma
+\frac{8}{3}a\alpha ^{3} -\frac{1}{3}a \beta -\alpha ^{2}\beta
-3\alpha ^{5}-\frac{4}{9}a^2\alpha)t^{2}+\cdots), \nonumber
\end{eqnarray}
with $\varepsilon=\pm i.$ Using the majorant method, we can show
that the formal Laurent series solutions are convergent.
Substituting the solutions $(7)$ into $F_{1}=c_{1},$ $F_{2}=c_{2}$
and $F_{3}=c_{3},$ and equating the $t^0$-terms yields
\begin{eqnarray}
F_{1}&=&\frac{15}{8}\alpha^{4}-\frac{5}{6}a\alpha^{2}-\frac{5}{4}\alpha
\beta -\frac{7}{36}a^{2}-\frac{5}{4}\varepsilon \sqrt{2}\theta=c_1, \nonumber\\
F_{2}&=&\varepsilon \sqrt{2}( \frac{1}{4}\alpha ^{5}-\gamma
+\frac{\varepsilon \sqrt{2}}{2}\alpha \theta -\frac{2}{3}a\alpha
^{3}+\frac{1}{3}a\beta +\frac{1}{6}a^{2}+\frac{1}{2}\alpha ^{2}\beta)=c_2,\nonumber\\
F_{3}&=&-\frac{11}{2}\alpha ^{6}-\beta ^{2}+4\alpha \gamma
+3\alpha ^{2}\varepsilon \sqrt{2}\theta +\alpha ^{3}\beta
-\frac{1}{3}a^{2}\alpha^{2}+\allowbreak\frac{10}{3}a\alpha^{4}=c_3.\nonumber
\end{eqnarray}
Eliminating $\gamma $ and $\theta$ from these equations, leads to
an equation connecting the two remaining parameters $\alpha$ and
$\beta$ :
\begin{equation}\label{eqn:euler}
\beta ^{2}+\frac{2}{3}(3\alpha ^{2}-2a)\alpha \beta -3\alpha ^{6}+
\frac{8}{3}a\alpha ^{4}+\frac{4}{9}(a^{2}+9c_{1})\alpha
^{2}-2\varepsilon \sqrt{2}c_{2}\alpha +c_{3}=0,
\end{equation}
According to Hurwitz' formula, this defines two isomorphic smooth
hyperelliptic curves $\mathcal{H}_\varepsilon$ ($\varepsilon =\pm
i$) of genus 2, which finishes the proof of the proposition.

\section{Affine part of an abelian surface as the jacobian of a
genus two hyperelliptic curve}

In order to embed $\mathcal{H}_{\varepsilon}$ into some projective
space, one of the key underlying principles used is the Kodaira
embedding theorem, which states that a smooth complex manifold can
be smoothly embedded into projective space $\mathbb{P}^N$ with the
set of functions having a pole of order k along positive divisor
on the manifold, provided k is large enough; fortunately, for
abelian varieties, k need not be larger than three according to
Lefshetz. These functions are easily constructed from the Laurent
solutions (7) by looking for polynomials in the phase variables
which in the expansions have at most a k-fold pole. The nature of
the expansions and some algebraic proprieties of abelian varieties
provide a recipe for when to terminate our search for such
functions, thus making the procedure implementable. Precisely, we
wish to find a set of polynomial functions $\{f_0,\ldots,f_N\},$
of increasing degree in the original variables $z_1,\ldots,z_5,$
having the property that the embedding $\mathcal{D}$ of
$\mathcal{H}_i+\mathcal{H}_{-i}$ into $\mathbb{P}^N$ via those
functions satisfies the relation : $$\mbox{geometric genus}
(2\mathcal{D})\equiv g(2\mathcal{D})=N+2.$$ A this point, it may
be not so clear why the curve $\mathcal{D}$ must really live on an
abelian surface. Let us say, for the moment, that the equations of
the divisor $\mathcal{D}$ (i.e., the place where the solutions
blow up), as a curve traced on the abelian surface $\widetilde{B}$
(to be constructed in proposition 4.2), must be understood as
relations connecting the free parameters as they appear firstly in
the expansions (7). In the present situation, this means that (8)
must be understood as relations connecting $\alpha$ and $\beta.$
Let
$$L^{(r)}=\left\{\begin{array}{rl}
&\mbox{polynomials}\quad f=f(z_,\ldots,z_5)\\& \mbox{of degre}
\leq r, \quad\mbox{with at worst a}\\& \mbox{double pole along}
\quad\mathcal{H}_i+\mathcal{H}_{-i}\\& \mbox{and with}\quad
z_1,\ldots,z_5 \quad\mbox{as in} (7)
\end{array}\right\}/[F_k=c_k, k=1,2,3],$$ and let
$(f_0,f_1,\ldots,f_{N_r})$ be a basis of $L^{(r)}.$ We look for r
such that :
$$g(2\mathcal{D}^{(r)})=N_r+2,\quad 2\mathcal{D}^{(r)}\subset
\mathbb{P}^{N_r}.$$ We shall show (proposition 4.1) that it is
unnecessary to go beyond r=4.

\begin{Lem}
The spaces $L^{(r)}$, nested according to weighted degree, are
generated as follows
\begin{eqnarray}
L^{(1)}&=&\{f_0,f_1,f_2,f_3,f_4,f_5\},\nonumber\\
L^{(2)}&=&L^{(1)}\oplus\{f_6,f_8,f_9,f_{10},f_{11},f_{12}\},\nonumber\\
L^{(3)}&=&L^{(2)},\nonumber\\
L^{(4)}&=&L^{(3)}\oplus\{f_{13},f_{14},f_{15}\},
\end{eqnarray}
where
$$f_0=1,\qquad f_1=z_1=\frac{\alpha}{t}+\ldots,$$
$$f_2=z_2=\frac{\varepsilon\sqrt{2}}{t}+\ldots,\qquad
f_3=z_3=-\frac{\varepsilon\sqrt{2}}{4t^2}+\ldots,$$
$$f_4=z_4=-\frac{\alpha}{2t^2}+\ldots,\qquad
f_5=z_5=-\frac{\eta}{3t}+\ldots, $$ $$f_6=z_{1}^{2}=\frac{\alpha
^{2}}{t^{2}}+\cdots,\qquad
f_7=z_{2}^{2}=-\frac{1}{8t^{2}}+\cdots,$$$$
f_8=z_{5}^{2}=\frac{\eta ^2}{9t^{2}}+\cdots,\qquad
f_9=z_{1}z_{2}=\frac{\varepsilon \sqrt{2}\alpha
}{4t^{2}}+\cdots,$$$$ f_{10}=z_{1}z_{5}=-\frac{\alpha
\eta}{3t^{2}} +\cdots,\qquad f_{11}=z_{2}z_{5}=-\frac{\varepsilon
\sqrt{2}\eta}{12t^{2}} +\cdots,$$$$ f_{12}=[ z_{1},z_{2}]
=-\frac{\varepsilon \sqrt{2}\alpha ^{2}}{2t^{2}}+\cdots,\qquad
f_{13}=[ z_{1},z_{5}] =\frac{4\alpha ^{2}\eta}{3t^{2}}
+\cdots,$$$$ f_{14}=[z_{2},z_{5}]=\frac{\varepsilon\sqrt{2}\alpha
\eta}{6t^{2}}+\cdots,\qquad f_{15}=(z_{3}-2\varepsilon
\sqrt{2}z_{2}^{2})^{2}=-\frac{\alpha ^{2}}{2t^{2}}+\cdots,$$ with
$$[z_j,z_k]=\dot z_j z_k-z_j\dot z_k,$$ the wronskien of $z_k$ and
$z_j,$ and $$\eta \equiv 3\beta -3\alpha ^{3}+a\alpha.$$
\end{Lem}
\emph{Proof}. The proof of this lemma is straightforward and can
be done by inspection of the expansions (7). Note also that the
functions $z_1,z_2,z_5$ behave as $1/t$ and if we consider the
derivatives of the ratios $z_1/z_2,$ $z_1/z_5,$ $z_2/z_5,$ the
wronskiens $[z_1,z_2],$ $[z_1,z_5],$ $[z_2,z_5],$  must behave as
$1/t^2$ since $z_2^2,z_5^2$ behave as $1/t^2$. This finishes the
proof of the lemma.

Note that $$dim L^{(1)}=6,\quad dim L^{(2)}=dim L^{(3)}=13,\quad
dim L^{(4)}=16.$$

\begin{Prop}
$L^{(4)}$ provides an embedding of $\mathcal{D}^{(4)}$ into
projective space $\mathbb{P}^{15}$ and $\mathcal{D}^{(4)}$ (resp.
$2\mathcal{D}^{(4)}$) has genus 5 (resp. 17).
\end{Prop}
\emph{Proof}. It turns out that neither $L^{(1)},$ nor $L^{(2)},$
nor $L^{(3)},$ yields a curve of the right genus; in fact
$$ g(2\mathcal{D}^{(r)})\neq \dim L^{(r)}+1,\quad r=1,2,3.$$
For instance, the embedding into $\mathbb{P}^{5}$ via $L^{(1)}$
does not separate the sheets, so we proceed to $L^{(2)}$ and we
consider the corresponding embedding into $\mathbb{P}^{12}.$ For
finite values of $\alpha$ and $\beta,$ the curves $\mathcal{H}_i$
and $\mathcal{H}_{-i}$ are disjoint; dividing the vector
$(f_0,\ldots,f_{12})$ by $f_7$ and taking the limit $t\rightarrow
0,$ to yield
$$[0:0:0:2\varepsilon \sqrt{2}:4\alpha:0:-8\alpha ^2:1:
-\frac{8}{9}\eta ^2:-2\varepsilon\sqrt{2}\alpha:\frac{8}{3}\alpha
\eta:\frac{2\varepsilon \sqrt{2}}{3}\eta:4\varepsilon
\sqrt{2}\alpha ^2].$$ The curve (8) has two points covering
$\alpha=\infty,$ at which $\eta \equiv 3\beta -3\alpha
^{3}+a\alpha $ behaves as follows :
\begin{eqnarray}
\eta &=&-6\alpha^3+3a\alpha\pm
3\sqrt{4\alpha^6-4a\alpha^4-4c_1\alpha^2+2\varepsilon\sqrt{2}c_2\alpha-c_3},\nonumber\\
&=&\left\{\begin{array}{rl}
-\frac{3(a^2+4c_1)}{4\alpha}+&\mbox{lower order terms},
\quad\mbox{picking the + sign},\\
-12\alpha^3+&O(\alpha), \quad\mbox{picking the - sign}.
\end{array}\right.\nonumber
\end{eqnarray}
Then by picking the - sign and by dividing the vector
$(f_0,\ldots,f_{12})$ by $f_8,$ the corresponding point is mapped
into the point $$[0:0:0:0:0:0:0:0:1:0:0:0:0],$$ in
$\mathbb{P}^{12}$ which is independent of $\varepsilon,$ whereas
picking the + sign leads to two different points, according to the
sign of $\varepsilon.$ Thus, adding at least 2 to the genus of
each curve, so that
$$g(2\mathcal{D}^{(2)})-2>12, \qquad 2\mathcal{D}^{(2)}\subset
\mathbb{P}^{12}\neq \mathbb{P}^{g-2},$$ which contradicts the fact
that $$N_r=g(2\mathcal{D}^{(2)})-2.$$ The embedding via $L^{(2)}$
(or $L^{(3)}$) is unacceptable as well. Consider now the embedding
$2\mathcal{D}^{(4)}$ into $\mathbb{P}^{15}$ using the 16 functions
$f_0,\ldots,f_{15}$ of $L^{(4)}$(9). It is easily seen that these
functions separate all points of the curve (except perhaps for the
points at $\infty$) : The curves $\mathcal{H}_i$ and
$\mathcal{H}_{-i}$ are disjoint for finite values of $\alpha$ and
$\beta$; dividing the vector $(f_0,\ldots,f_{15})$ by $f_7$ and
taking the limit $t\rightarrow 0,$ to yield
$$[0:0:0:2\varepsilon \sqrt{2}:4\alpha:0:-8\alpha ^2:1:
-\frac{8}{9}\eta ^2:-2\varepsilon\sqrt{2}\alpha:\frac{8}{3}\alpha
\eta:\frac{2\varepsilon \sqrt{2}}{3}\eta:$$$$ 4\varepsilon
\sqrt{2}\alpha ^2:-\frac{32}{3}\alpha^2\eta:-\frac{4\varepsilon
\sqrt{2}}{3}\alpha \eta:4\alpha^2].$$ About the point
$\alpha=\infty,$ it is appropriate to divide by $g_8;$ then by
picking the sign - in $\eta$ above, the corresponding point is
mapped into the point
$$[0:0:0:0:0:0:0:0:1:0:0:0:0:0:0:0],$$ in $\mathbb{P}^{15}$
which is independent of $\varepsilon,$ whereas picking the + sign
leads to two different points, according to the sign of
$\varepsilon.$ Hence, the divisor $\mathcal{D}^{(4)}$ obtained in
this way has genus 5 and thus $g(2\mathcal{D}^{(4)})$ has genus 17
and $$2\mathcal{D}^{(4)}\subset
\mathbb{P}^{15}=\mathbb{P}^{g-2},$$ as desired. This ends the
proof of the proposition.

Let $$L=L^{(4)},\quad\mathcal{D}=\mathcal{D}^{(4)},$$ and
$$\mathcal{S}=2\mathcal{D}^{(4)}\subset \mathbb{P}^{15}.$$ Next we
wish to construct a surface strip around $\mathcal{S}$ which will
support the commuting vector fields. In fact, $\mathcal{S}$ has a
good chance to be very ample divisor on an abelian surface, still
to be constructed.

\begin{Prop}
The variety B (3) generically is the affine part of an abelian
surface $\widetilde{B},$ more precisely the jacobian of a genus 2
curve. The reduced divisor at infinity
$$\widetilde{B}\backslash B=\mathcal{H}_i+\mathcal{H}_{-i},$$
consists of two smooth isomorphic genus 2 curves
$\mathcal{H}_\varepsilon $(8). The system of differential
equations (1) is algebraic complete integrable and the
corresponding flows evolve on $\widetilde{B}.$
\end{Prop}
\emph{Proof}. We need to attaches the affine part of the
intersection of the three invariants (2) so as to obtain a smooth
compact connected surface in $\mathbb{P}^{15}$. To be precise, the
orbits of the vector field (1) running through $\mathcal{S}$ form
a smooth surface $\Sigma$ near $\mathcal{S}$\ such that
$$\Sigma \backslash B \subseteq \widetilde{B},$$ and the variety
$$\widetilde{B}=B\cup \Sigma,$$ is smooth, compact and connected.
Indeed, let $$\psi(t,p)=\{ z(t)=(z_{1}(t),\ldots,z_{5}(t)):t\in
\mathbb{C},0< |t| < \varepsilon \},$$ be the orbit of the vector
field (1) going through the point $p\in \mathcal{S}$. Let $\Sigma
_{p}\subset \mathbb{P}^{15}$ be the surface element formed by the
divisor $\mathcal{S}$\ and the orbits going through $p$. Consider
the curve $$\mathcal{S}^{\prime }=\mathcal{H}\cap \Sigma ,$$ where
$\mathcal{H}\subset \mathbb{P}^{15}$ is a hyperplane transversal
to the direction of the flow and $\Sigma \equiv
\displaystyle{\cup_{p\in \mathcal{S}}\Sigma _{p}}$. If
$\mathcal{S}^{\prime }$\ is smooth, then using the implicit
function theorem the surface $\Sigma $\ is smooth. But if
$\mathcal{S}^{\prime }$\ is singular at $0,$\ then $\Sigma $\
would be singular along the trajectory ($t-$axis) which go
immediately into the affine part B. Hence, B would be singular
which is a contradiction because B is the fibre of a morphism from
$\mathbb{C}^{5}$\ to $\mathbb{C}^{2}$\ and so smooth for almost
all the three constants of the motion $c_{k}.$ Next, let
$\overline{B}$ be the projective closure of B into
$\mathbb{P}^{5},$ let $Z=[Z_{0}:Z_{1}:\ldots:Z_{5}]\in
\mathbb{P}^{5}$ and let $I=\overline{B}\cap \{Z_0=0\}$ be the
locus at infinity. Consider the map $$\overline{B}\subseteq
\mathbb{P}^{5} \longrightarrow \mathbb{P}^{15},\text{
}Z\longmapsto f(Z),$$ where $f=(f_0,f_{1},...,f_{15}) \in
L(\mathcal{S})$(9) and let $\widetilde{B}=f(\overline{B}).$ In a
neighbourhood $V(p)\subseteq \mathbb{P}^{15}$ of $p$, we have
$\Sigma _{p}= \widetilde{B}$ and $\Sigma _{p}\backslash
\mathcal{S}\subseteq B$. Otherwise there would exist an element of
surface $\Sigma _{p}^{\prime }\subseteq \widetilde{B}$ such that
$$\Sigma _{p}\cap \Sigma _{p}^{\prime }=(t-axis), \mbox{orbit}\psi
(t,p)=(t-axis)\backslash \ p\subseteq B,$$ and hence B would be
singular along the $t-$axis which is impossible. Since the variety
$\overline{B}\cap \{Z_{0}\neq 0\}$ is irreducible and since the
generic hyperplane section $\mathcal{H}_{gen.}$\ of $\overline{B}$
is also irreducible, all hyperplane sections are connected and
hence I is also connected. Now, consider the graph $\Gamma
_{f}\subseteq \mathbb{P}^{5}\times \mathbb{P}^{15}$ of the map
$f,$ which is irreducible together with $\overline{B}.$\ It
follows from the irreducibility of I that a generic hyperplane
section $\Gamma _{f}\cap \{\mathcal{H}_{gen.}\times
\mathbb{P}^{15}\}$ is irreducible, hence the special hyperplane
section $\Gamma _{f}\cap \{\{Z_{0}=0\}\times \mathbb{P}^{15}\}$ is
connected and therefore the projection map
$$proj_{\mathbb{P}^{15}}\{\Gamma _{f}\cap \{\{Z_{0}=0\}\times
\mathbb{P}^{15}\}\}=f(I)\equiv \mathcal{S},$$ is connected. Hence,
the variety $B\cup \Sigma =\widetilde{B}$ is compact, connected
and embeds smoothly into $\mathbb{P}^{15}$ via $f.$ We wish to
show that $\widetilde{B}$ is an abelian surface equipped with two
everywhere independent commuting vector fields. For doing that,
let $\phi^{\tau _{1}}$\ and $\phi^{\tau _{2}}$\ be the flows
corresponding to vector fields $X_{F_{1}}$ and $X_{F_{2}}$. The
latter are generated respectively by $F_1$ and $F_2.$ For $p\in
\mathcal{S}$ and for small $\varepsilon
> 0,$ $$\phi^{\tau _{1}}(p), \quad\forall \tau _{1},0<|\tau _{1}|<
\varepsilon ,$$ is well defined and $\phi^{\tau _{1}}(p)\in B$.
Then we may define $\phi^{\tau _{2}}$ on B by
$$
\phi^{\tau _{2}}(q)=\phi^{-\tau _{1}}\phi^{\tau _{2}}\phi^{\tau
_{1}}(q),\quad q\in U(p)=\phi^{-\tau _{1}}(U(\phi^{\tau
_{1}}(p))),
$$
where $U(p)$ is a neighbourhood of $p$. By commutativity one can
see that $\phi^{\tau _{2}}$ is independent of $\tau _{1};$
\begin{eqnarray}
\phi^{-\tau _{1}-\varepsilon _{1}}\phi^{\tau _{2}}\phi^{\tau
_{1}+\varepsilon _{1}}(q)&=&\phi^{-\tau _{1}}\phi^{-\varepsilon
_{1}}\phi^{\tau _{2}}\phi^{\tau _{1}}\phi^{\varepsilon
_{1}},\nonumber\\
&=&\phi^{-\tau _{1}}\phi^{\tau _{2}}\phi^{\tau _{1}}(q).\nonumber
\end{eqnarray}
We affirm that $\phi^{\tau _{2}}(q)$ is holomorphic away from
$\mathcal{S}.$ This because $\phi^{\tau _{2}}\phi^{\tau _{1}}(q)$
is holomorphic away from $\mathcal{S}$ and that $\phi^{\tau _{1}}$
is holomorphic in $U(p)$ and maps bi-holomorphically $U(p)$ onto
$U(\phi^{\tau _{1}}(p)).$ Now, since the flows $\phi^{\tau _{1}}$
and $\phi^{\tau _{2}}$ are holomorphic and independent on
$\mathcal{S},$ we can show along the same lines as in the
Arnold-Liouville theorem [1,6] that $\widetilde{B}$ is a complex
torus $\mathbb{C}^{2}/lattice$ and so in particular
$\widetilde{B}$ is a K\"{a}hler variety$.$ And that will done, by
considering the local diffeomorphism
$$\mathbb{C}^{2}\longrightarrow \widetilde{B}, (\tau _{1},\tau
_{2})\longmapsto \phi^{\tau _{1}}\phi^{\tau _{2}}(p),$$ for a
fixed origin $p\in B.$ The additive subgroup $$\{(\tau _{1},\tau
_{2})\in \mathbb{C}^{2}:\phi^{\tau _{1}}\phi^{\tau _{2}}(p)=p\},$$
is a lattice of $\mathbb{C}^{2}$, hence
$$\mathbb{C}^{2}/lattice\longrightarrow \widetilde{B},$$ is a
biholomorphic diffeomorphism and $\widetilde{B}$ is a K\"{a}hler
variety with K\"{a}hler metric given by $d\tau _{1}\otimes
d\overline{\tau }_{1}+d\tau _{2}\otimes d\overline{\tau }_{2}.$
Now, a compact complex K\"{a}hler variety having the required
number as (its dimension) of independent meromorphic functions is
a projective variety [11]. In fact, here we have
$\widetilde{B}\subseteq \mathbb{P}^{15}.$ Thus $\widetilde{B}$ is
both a projective variety and a complex torus
$\mathbb{C}^{2}/lattice$ and hence an abelian surface as a
consequence of Chow theorem. By the classification theory of ample
line bundles on abelian varieties, $\widetilde{B}\simeq
\mathbb{C}^{2}/L_{\Omega}$ with period lattice given by the
columns of the matrix
$$ \left(\begin{array}{cccc}
\delta_1&0&a&c\\ 0&\delta_2&c&b
\end {array}\right),\qquad
Im\left(\begin{array}{cc} a&c\\ c&b
\end {array}\right)>0,$$
and $$\delta_1 \delta_2=g(\mathcal{H}_\varepsilon)-1=1,$$ implying
$\delta_1=\delta_2=1.$ Thus $\widetilde{B}$ is principally
polarized and it is the jacobian of the hyperelliptic curve
$\mathcal{H}_\varepsilon.$ This completes the proof of the
proposition.

\begin{Rem}
We have seen that the reflection $\sigma$ on the affine variety B
amounts to the flip $$\sigma :(z_1,z_2,z_3,z_4,z_5)\longmapsto
(z_1,z_2,-z_3,-z_4,z_5),$$ changing the direction of the commuting
vector fields. It can be extended to the (-Id)-involution about
the origin of $\mathbb{C}^2$ to the time flip $(t_1,t_2)\mapsto
(-t_1,-t_2)$ on $\widetilde{B}$, where $t_{1}$ and $t_{2}$ are the
time coordinates of each of the flows $X_{{F}_{1}}$ and
$X_{{F}_{2}}.$ The involution $\sigma $ acts on the parameters of
the Laurent solution (7) as follows
$$\sigma :(t,\alpha,\beta,\gamma,\theta,\varepsilon)\longmapsto
(-t,-\alpha,-\beta,-\gamma,-\theta,-\varepsilon),$$ interchanges
the curves $\mathcal{H}_{\varepsilon =\pm i}$ (8) and the linear
space $L$ can be split into a direct sum of even and odd
functions. Geometrically, this involution interchanges
$\mathcal{H}_{i}$ and $\mathcal{H}_{-i},$ i.e.,
$\mathcal{H}_{-i}=\sigma \mathcal{H}_{i}.$
\end{Rem}

\begin{Rem}
Consider on $\widetilde{B}$ the holomorphic 1-forms $dt_1$ and
$dt_2$ defined by $dt_i(X_{F_j})=\delta_{ij},$ where $X_{F_1}$ and
$X_{F_2}$ are the vector fields generated respectively by $F_1$
and $F_2.$ Taking the differentials of $\zeta=1/z_1$ and
$\xi=z_1/z_2$ viewed as functions of $t_1$ and $t_2,$ using the
vector fields and the Laurent series (7) and solving linearly for
$dt_1$ and $dt_2,$ we obtain as expected the hyperelliptic
holomorphic differentials
\begin{eqnarray}
\omega_1&=&dt_{1}|_{\mathcal{H}_{\varepsilon}},\nonumber\\
&=&\frac{1}{\triangle}(\frac{\partial \xi}{\partial
t_2}d\zeta-\frac{\partial \zeta}{\partial
t_2}d\xi)|_{\mathcal{H}_{\varepsilon}},\nonumber\\
&=&\frac{\alpha
d\alpha}{\sqrt{P(\alpha)}},\nonumber\\
\omega_2&=&dt_{2}|_{\mathcal{H}_{\varepsilon}},\nonumber\\
&=&\frac{1}{\triangle}(\frac{-\partial \xi}{\partial
t_1}d\zeta-\frac{\partial \zeta}{\partial
t_1}d\xi)|_{\mathcal{H}_{\varepsilon}},\nonumber\\
&=&\frac{\sqrt{2}d\alpha}{2\sqrt{P(\alpha)}},\nonumber
\end{eqnarray}
with $$P(\alpha)\equiv 4\alpha ^{6}-4a\alpha ^{4}-4c_{1}\alpha
^{2}+2\varepsilon \sqrt{2}c_{2}\alpha -c_{3},$$ and $$\Delta
\equiv {\frac{\partial \zeta}{\partial t_1}\frac{\partial
\xi}{\partial t_2}-\frac{\partial \zeta}{\partial
t_2}\frac{\partial \xi}{\partial t_1}}.$$ The zeroes of $\omega_2$
provide the points of tangency of the vector field $X_{F_{1}}$ to
$\mathcal{H}_\varepsilon.$ We have
$\frac{\omega_1}{\omega_2}=-\varepsilon \sqrt{2}\alpha,$ and
$X_{F_{1}}$ is (doubly) tangent to $\mathcal{H}_\varepsilon$ at
the point covering $\alpha =\infty,$ i.e., where both the curves
touch.
\end{Rem}

\end{document}